\documentclass[preprint, showpacs, preprintnumbers,amsmath,amssymb,nofootinbib]{revtex4}
\usepackage{mathrsfs}
\usepackage{amsmath}
\usepackage[colorlinks, linkcolor=blue, citecolor=red, urlcolor=red]{hyperref}
\usepackage{graphicx}
\usepackage{bm}
\usepackage{psfrag}
\usepackage{amsmath}
\usepackage{amssymb}
\usepackage{latexsym}
\usepackage{exscale}
 \begin{document}
\title{Determining $H_0$ using a model-independent method}
\author{Puxun Wu$^{1,2}$, Zhengxiang Li$^3$, and Hongwei Yu$^1$}
\address{ $^1$Center for Nonlinear Science and Department of Physics, Ningbo
University, Ningbo, Zhejiang 315211, China\\
$^2$Center for High Energy Physics, Peking University, Beijing 100080, China\\
$^3$Department of Astronomy, Beijing Normal University, Beijing 100875, China}
\begin{abstract}
 By using type Ia supernovae (SNIa) to provide the luminosity distance (LD) directly, which depends on the value of the Hubble constant $H_0= 100 h\; {\rm km\; s^{-1}\; Mpc^{-1}}$, and the angular diameter distance from galaxy clusters or baryon acoustic oscillations (BAOs) to give the derived LD according to the distance duality relation, we propose a model-independent method to determine $h$ from  the fact that different observations should give the same LD at a given redshift. Combining the Sloan Digital Sky Survey II (SDSS-II) SNIa from the MLCS2k2 light curve fit and galaxy cluster data, we find that at the $1\sigma$ confidence level (CL), $h=0.5867\pm0.0303$ for the sample of the elliptical $\beta$ model for galaxy clusters, and $h=0.6199\pm0.0293$ for that of the spherical $\beta$ model. The former is smaller than the values from other observations, whereas the latter is consistent with the Planck result at the $2\sigma$ CL and agrees very well with the value reconstructed directly from the $H(z)$ data. With the SDSS-II SNIa and BAO measurements, a tighter constraint, $h=0.6683\pm0.0221$, is obtained. For comparison, we also consider the Union 2.1 SNIa from the SALT2 light curve fitting. The results from the Union 2.1 SNIa are slightly larger than those from the SDSS-II SNIa, and the Union 2.1 SNIa + BAOs give the tightest value. We find that the values from SNIa + BAOs are quite consistent with those from the Planck and the BAOs, as well as the local measurement from Cepheids and very-low-redshift SNIa.
 
\end{abstract}
\maketitle
\section{Introduction}
The determination of the Hubble constant $H_0$ from astronomical observations is very important in cosmology, as it gives the present cosmic expansion speed and is related to the cosmic components and the size and age of our Universe. Its first accurate value, $H_0=72\pm 8 \; {\rm km\; s^{-1}\; Mpc^{-1}}$ at the $1\sigma$ confidence level (CL)~\cite{Freedman2001}, was obtained using local measurements from the Hubble Space Telescope. About 10 years later, Riess et al.~\cite{Riess2011} improved this result significantly using Cepheids and very-low-redshift Type Ia supernovae (SNIa) and obtained $H_0=73.8\pm 2.4 \; {\rm km\; s^{-1}\; Mpc^{-1}}$, where the uncertainty was reduced from $\sim11\%$ to $\sim3\%$. In addition to the cosmic distance ladder, the Hubble constant can also be determined using cosmic microwave background (CMB) measurement. On the basis of a six-parameter $\Lambda$ cold dark matter (CDM) model, the nine-year Wilkinson Microwave Anisotropy Probe (WMAP9) data give $H_0=70.0\pm 2.2 \; {\rm km\; s^{-1}\; Mpc^{-1}}$, a $3\%$ determination~\cite{Bennett2013, Hinshaw2013}. Combining WMAP9 with smaller angular scale CMB data from the South Pole Telescope (SPT) and the Atacama Cosmology Telescope (ACT) as well as distance measurements derived from baryon acoustic oscillation (BAO) observations improves this determination to $1.2\%$ and gives $H_0=68.76\pm 0.84 \; {\rm km\; s^{-1}\; Mpc^{-1}}$~\cite{Bennett2014}, which is consistent with the result ($H_0=68\pm 2 \; {\rm km\; s^{-1}\; Mpc^{-1}}$) determined by median statistics from Huchra's final compilation~\cite{Chen11, Calabrese12}. Recently, in the framework of a six-parameter $\Lambda$CDM model, the Planck Collaboration also found that $H_0=67.3\pm 1.2 \; {\rm km\; s^{-1}\; Mpc^{-1}}$~\cite{Planck} and $H_0=67.8\pm 0.9 \; {\rm km\; s^{-1}\; Mpc^{-1}}$~\cite{Planck XIII}, which are consistent with the value ($H_0=67.3\pm 1.1 \; {\rm km\; s^{-1}\; Mpc^{-1}}$)~\cite{1411} obtained by combining galaxy BAO measurements from the Baryon Oscillation Spectroscopic Survey (BOSS) survey of Sloan Digital Sky Survey III (SDSS-III)~\cite{Anderson13} with high-precision measurements of the relative distances of the latest SNIa~\cite{SN}. However, the Planck results are slightly smaller than the WMAP9 one and have a tension of about $2.5 \sigma $ with respect to local measurements given in~\cite{Riess2011}. 

On one hand, this tension might indicate the need for new physics such as the cosmic variance~\cite{Marra} or a super-accelerating expansion~\cite{Zhang}, but on the other hand, it may also be due to systematic errors~\cite{Efstathiou2014, Rigault14, Spergel 2013, Romanoa, Romanob}. For example, revising the geometric maser distance to NGC 4258 from \cite{Humphreys2013} and using this indicator to calibrate the data considered in~\cite{Riess2011}, Efstathiou~\cite{Efstathiou2014} yielded $H_0=70.6\pm 3.3 \; {\rm km\; s^{-1}\; Mpc^{-1}}$. Considering predominantly star-forming environments, Rigault et al.~\cite{Rigault14} found $H_0=70.6\pm 2.6 \; {\rm km\; s^{-1}\; Mpc^{-1}}$. These results are consistent with that of Planck. In addition, the systematic errors in the CMB analysis may also be responsible for some part of the tension. By removing the $217\times 217 \; {\rm GHz}$ detector set spectrum used in the Planck analysis, it was found that $H_0=68.0\pm 1.1 \; {\rm km\; s^{-1}\; Mpc^{-1}}$~\cite{Spergel 2013}.

In addition to the very-low-redshift data from local measurements and the CMB data at $z=1089$, the intermediate-redshift data provide a complementary tool for determining the Hubble constant, and they help to reduce the uncertainties of $H_0$ significantly if they are combined with CMB data. Cheng and Huang~\cite{Cheng2014} combined BAO data from the 6dF Galaxy Survey~\cite{Beutler11}, BOSS Data Release 11 (DR11) clustering of galaxies~\cite{Anderson13}, WiggleZ~\cite{Kazin14}, and $z = 2.34$ from the BOSS DR11 quasar Lyman-$\alpha$ forest lines~\cite{Delubac14}, as well as simultaneous measurements of the two-dimensional two-point correlation function from the BOSS DR9 CMASS sample~\cite{Chuang13} and the two-dimensional matter power spectrum from the SDSS DR7 sample~\cite{Hemantha13}, and found that in a $\Lambda$CDM model, $H_0=68.0\pm 1.1 \; {\rm km\; s^{-1}\; Mpc^{-1}}$, a $1.3\%$ determination. Using the WMAP9+SPT+ACT+6dFGS+BOSS/DR11+$H_0$/Riess and basing their analysis on the six-parameter $\Lambda$CDM cosmology, Bennett et al.~\cite{Bennett2014} obtained $H_0=69.6\pm 0.7 \; {\rm km\; s^{-1}\; Mpc^{-1}}$, a $1\%$ determination, which is the most accurate result to date. Through a nonparametric reconstruction of $H(z)$ data, Busti et al.~\cite{Busti2014} obtained $H_0=64.9\pm 4.2 \; {\rm km\; s^{-1}\; Mpc^{-1}}$ by extrapolating the reconstruction to redshift $0$. All these results seem to be consistent with that of Planck. However, using other intermediate-redshift data including angular diameter distances (ADDs) from the spherical $\beta$ model galaxy cluster sample~\cite{Bonamente2006}, 11 ages of old high-redshift galaxies~\cite{Ferreras09, Longhetti07}, 18 $H(z)$ data points~\cite{Simon05, Gaztanaga09, Stern10}, and the BAO peak at $z=0.35$~\cite{Eisenstein05}, Lima and Cunha~\cite{Lima2014} found $H_0=74.1\pm 2.2 \; {\rm km\; s^{-1}\; Mpc^{-1}}$ in a $\Lambda$CDM model. This value is consistent with the result obtained in~\cite{Riess2011}. Later, Holanda et al.~\cite{Holanda2014} considered the effects of different galaxy cluster samples and found that different samples give different results, i.e., $H_0=70\pm 4 \; {\rm km\; s^{-1}\; Mpc^{-1}}$ if the spherical $\beta$ model galaxy cluster sample~\cite{Bonamente2006} is replaced by the elliptical one~\cite{Filippis2005}. Therefore, not only local and global measurements but also different intermediate-redshift data may yield controversial results for the value of the Hubble constant.

The above discussion illustrates that, except for direct measurement from the cosmic distance ladder and nonparametric reconstruction from $H(z)$ data, the reported results are based on the $\Lambda$CDM model. In this paper, we propose a new model-independent method to determine $H_0$ by combining the observed luminosity distances (LDs) from SNIa data and ADDs from galaxy clusters or BAOs.
\section{Method}
The measurement of SNIa is very important in modern cosmology, and it first indicates that the present expansion of our Universe is speeding up. The observed SNIa data are usually released in the form of the distance modulus $\mu$, which is related to the LD $D_{\rm L}$ by
\begin{eqnarray}\label{mu0}
\mu=5\log_{10} \frac{ D_{\rm L}}{{\rm Mpc}}+25\;.
\end{eqnarray}
The value of the distance modulus is determined mainly by the rest-frame peak magnitude $m$ and the peak absolute magnitude $M$ of SNIa. Because $M$ is degenerate with $H_0=100 h \; {\rm km\; s^{-1}\;Mpc^{-1}}$, we must choose a value of $h$ to obtain the distance modulus. For example, $h=0.65$ in the SDSS-II sample~\cite{Kessler09}, and $h=0.70$ in the Union 2 one~\cite{Amanullah10}. Because $D_{\rm L} \propto \frac{1}{h}$ in the Friedmann--Robertson--Walker universe, letting $h_0$ denote the chosen value of $h$ to determine the distance modulus, one finds that 
\begin{eqnarray}\label{mu4}
\mu=\mu_{0}-5\log_{10}\frac{h}{h_0}\;,
\end{eqnarray}
where $\mu_0$ represents the released distance modulus with $h=h_0$. Combining Eq.~(\ref{mu4}) with Eq.~(\ref{mu0}) 
 gives the $h$-dependent LD from the distance modulus of SNIa:
\begin{eqnarray}\label{DL}
 D_{\rm L}(h)=\frac{h_0}{h}10^{\mu_0/5-5}\; {\rm Mpc}\equiv \frac{h_0}{h}D_{\rm L0}\;.
\end{eqnarray}

Several light curve fitting methods are used to analyze SNIa data. Here, we consider the MLCS2k2~\cite{Jha07} fitting method and the SALT2~\cite{Guy07} method for comparison. In the MLCS2k2 fitter, the estimated value and uncertainty of the distance modulus are obtained independent of any cosmological models. For this fitter, we use the 288 SDSS-II SNIa data points~\cite{Kessler09} in our analysis, which are given with $h_0=0.65$. 

For the SALT2 light curve fitting method, the released distance modulus is model-dependent, as the observed value of the distance modulus is obtained using $\mu=m-M+\alpha x-\beta c$, 
where $x$ is a stretch factor that describes the effects of the shapes of light curves on $\mu$, and $c$ is a color parameter that denotes the influences of the intrinsic color and reddening by dust. $\alpha$, $\beta$, and $M$ are fitted by minimizing the residuals in the Hubble diagram in the framework of the $\Lambda$CDM model. For the SALT2 fit, we consider the Union 2.1~\cite{Suzuki12} SNIa sample, which is obtained with $h_0=0.7$ and consists of 580 data points.

From Eq.~(\ref{DL}), one can see that, if the LD can be obtained from other observations and is independent of $h$, the value of the Hubble constant can be determined directly by combining them with the SNIa observations, and the result is model-independent~\footnote{Notice that for the Union 2.1 data, the result still depends on the $\Lambda$CDM model.}. However, no observations other than SNIa can provide the observed value of the LD directly. Fortunately, the ADD $D_{\rm A}$ can be obtained from some observations, such as galaxy clusters and BAOs. Using the distance duality relation between the LD and the ADD, $D_{\rm L}=(1+z)^2 D_{\rm A}$, we can deduce the value of $D_{\rm L}$ at a redshift of $D_{\rm A}$. Thus, if one has observed SNIa and galaxy cluster (or BAO) data at the same redshift and assumes that the distance duality relation is valid, the combination of these data provides a model-independent method to constrain $h$. This assumption is reasonable because the distance duality relation is valid in any metric theory of gravity, and its validity has been tested in~\cite{Bassett03, Nair, Lampeitl, Kunz04, Li2011} using different datasets.
 
The above analysis shows that finding the observed LD ($D_{\rm L0}$) from the distance modulus of SNIa and the ADD from galaxy clusters or BAOs at the same redshift is necessary to constrain $h$ in a model-independent way. Because the SNIa data sample is much larger than that of galaxy clusters or BAOs, to obtain a tight constraint, all ADD data should be kept. Owing to the usual absence of SNIa data at the same redshift as ADD data, we use a binning method to give the corresponding LD value of SNIa. This means that all SNIa data available in the range $\Delta z=|z_{\rm L}-z_{\rm A}|\leq \Delta$ are binned, where $z_{\rm A}$ represents the redshift of the ADD data, and constant $\Delta$ represents the binned redshift region. Using the criterion that the standard deviation of the set of $\mu_0 $ in a bin has the same order of magnitude as $\sigma_{\bar{\mu}_{\rm 0}}$, which is defined in Eq.~(\ref{sig}), to determine the value of $\Delta$, we choose $\Delta=0.011$. For all the selected data, an inverse variance weighted average is employed. If $\mu_{{\rm 0},m}$ denotes the $m$th appropriate SNIa distance modulus data in a bin with the given redshift $z_i$, in which there are $N_i$ SNIa data points, and $\sigma_{\mu_{{\rm 0},m}}$ represents the corresponding observational uncertainty, one can straightforwardly find, using the conventional data reduction techniques in~\cite{Bevington2003}, that 
\begin{equation}\label{bu1}
\bar{\mu}_{{\rm 0},i}={\sum_{m=1}^{N_i} (\mu_{{\rm 0},m}/\sigma_{\mu_{{\rm 0},m}}^2)\over \sum_{m=1}^{N_i} 1/\sigma_{\mu_{{\rm 0},m}}^2},
\end{equation}
\begin{equation}\label{sig}
\sigma^2_{\bar{\mu}_{{\rm 0},i}}={1\over \sum_{m=1}^{N_i} 1/\sigma_{\mu_{{\rm 0},m}}^2},
\end{equation}
where $\bar{\mu}_{{\rm 0},i}$ represents the weighted mean distance modulus
at a given redshift $z_i$, and $\sigma^2_{\bar{\mu}_{{\rm 0},i}}$ is its uncertainty. 

 If the observational uncertainties of the distance modulus are released with the covariance matrix $C$, using a method similar to that given in~\cite{Bevington2003}, we find that the weighted mean distance modulus in a bin and its uncertainty can be expressed as follows: 
\begin{eqnarray}\label{bu2}
\bar{\mu}_{{\rm 0},i}=\frac{\sum_{m,n=1}^{N_i}  \mu_{{\rm 0}, m}C_{mn}^{-1}}{\sum_{m,n=1}^{N_i} C_{mn}^{-1}},
\end{eqnarray}
\begin{equation}\label{sig2}
\sigma^2_{\bar{\mu}_{\rm 0}, i}={1\over \sum_{m,n=1}^{N_i}  C_{mn}^{-1}}.
\end{equation} 
When only the statistical errors are considered, $C$ is a diagonal matrix, and $C_{nn}=\sigma^2_{{\mu}_{{\rm 0},n}}$. In this case, Eqs.~(\ref{bu2}) and (\ref{sig2}) reduce to Eqs.~(\ref{bu1}) and (\ref{sig}), respectively.

Using Eq.~(\ref{mu0}), one can obtain the binned luminosity distance and the corresponding uncertainty: 
\begin{eqnarray}
\bar{D}_{{\rm L0},i}= 10^{\frac{\bar{\mu}_{{\rm 0},i}}{5}-5} {\rm Mpc},
\end{eqnarray}
\begin{eqnarray}
\sigma_{\bar{D}_{{\rm L0},i}}=\frac{\ln 10}{5} 10^{\frac{\bar{\mu}_{{\rm 0},i}}{5}-5} \sigma_{\bar{\mu}_{{\rm 0},i}}.
\end{eqnarray}
Then, from Eq.~(\ref{DL}) and the distance duality relation, the observed $h_{obs,i}$ can be obtained:
\begin{eqnarray}
 h_{obs,i}=\frac{h_0 \bar{D}_{{\rm L0},i}}{D_{{\rm A},i}}{(1+z_i)}^{-2}\;,
 \end{eqnarray}
with 
$\sigma_{h_{obs,i}}^2=h_0^2  D_{{\rm A},i}^{-2}{(1+z_i)}^{-4}  \sigma^2_{\bar{D}_{{\rm L0},i}} +h_0^2  \bar{D}_{{\rm L0},i}^2  D_{{\rm A},i}^{-4}{(1+z_i)}^{-4}  \sigma^2_{D_{{\rm A},i}}$. Here, $ D_{{\rm A},i}$ is the observed ADD at the redshift $z_i$, and $\sigma_{D_{{\rm A},i}}$ is its uncertainty. Minimizing the $\chi^2$ statistics
 \begin{eqnarray}
\chi^2=\sum_i \frac{(h- h_{obs,i})^2}{ \sigma_{h_{obs,i}}^2}\;,
 \end{eqnarray}
we can obtain the constraint on $h$. The best-fit value is given by $\chi^2_{min}$, which is the minimum of $\chi^2$. The $1\sigma$, $2\sigma$, and $3\sigma$ CLs are determined through $\Delta\chi^2=\chi^2-\chi^2_{min}\leq 1$, $\Delta\chi^2\leq 4$, and $\Delta\chi^2\leq 9$, respectively. The probability density of $h$, $P(h)=A\exp (-\chi^2/2)$, can be obtained, where $A$ is a normalized coefficient.
\section{galaxy cluster sample}
In \cite{Reese02}, it was suggested that the ADD of a galaxy cluster can be derived from a measurement of its intrinsic size on the basis of the Sunyaev--Zel'dovich (SZ) effect and X-ray surface brightness observations. Here, we employ two different galaxy cluster samples: the spherical $\beta$ model sample and elliptical $\beta$ model sample. The former, compiled in \cite{Bonamente2006}, has $38$ ADD data points in the redshift region $0.14<z<0.89$, which are obtained by using a hydrostatic equilibrium model to analyze the cluster plasma and dark matter distributions. The latter was compiled in \cite{Filippis2005} using an isothermal elliptical $\beta$ model and consists of $25$ ADD data in the redshift region $0.02<z<0.79$, in which 18 galaxy clusters were compiled in \cite{Reese02} and 7 galaxy clusters were compiled in \cite{Mason01}.
\subsection{Galaxy cluster sample + SDSS-II SNIa}
\begin{figure}
\includegraphics[width=8cm]{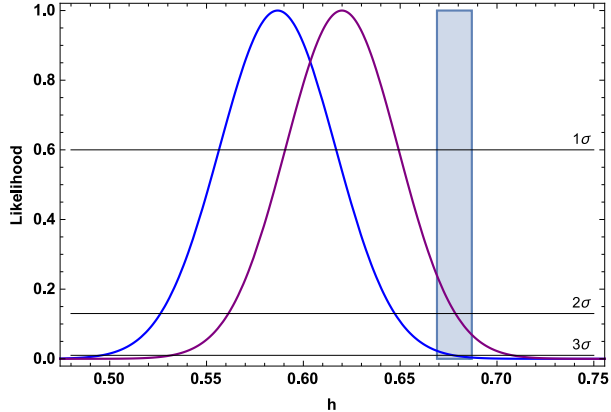} 
\caption{The likelihood distributions of $h$ from galaxy cluster + SDSS-II SNIa. The blue and purple solid lines represent the results from SDSS-II SNIa + 25 elliptical $\beta$ model sample, and SDSS-II SNIa + 38  spherical $\beta$ model sample, respectively. The shaded region shows the constraint from the latest Planck measurement with the $1\sigma$ CL ($H_0=67.8\pm 0.9 \; {\rm km\; s^{-1}\; Mpc^{-1}}$)~\cite{Planck XIII}. \label{fig2}}
\end{figure}

The systematic errors in the SDSS-II SNIa data sample are not considered in our analysis. The binned distance modulus can be obtained from Eqs.~(\ref{bu1}) and (\ref{sig}). Fig.~(\ref{fig2}) shows the likelihood distributions of $h$, where the blue and purple solid lines represent the results from the SDSS-II SNIa + 25 elliptical $\beta$ model sample and SDSS-II SNIa + 38 spherical $\beta$ model sample, respectively, and the shaded region shows the constraint from the latest Planck measurement with the $1\sigma$ CL ($H_0=67.8\pm 0.9 \; {\rm km\; s^{-1}\; Mpc^{-1}}$)~\cite{Planck XIII}. We find that at the $1\sigma$ CL, $h=0.5867\pm0.0303$ for the SDSS-II SNIa + 25 galaxy clusters, and $h=0.6199\pm0.0293$ for the SDSS-II SNIa + 38 galaxy clusters. 
\subsection{Galaxy cluster sample + Union 2.1 SNIa}
\begin{figure}
\includegraphics[width=8cm]{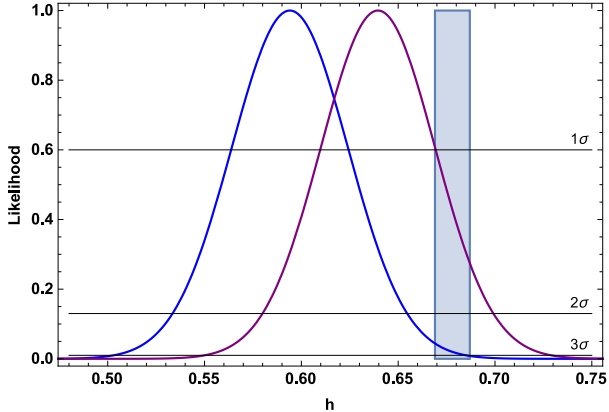} 
\caption{The likelihood distributions of $h$ from galaxy cluster + Union 2.1 SNIa. The blue and purple solid lines represent the results from Union 2.1 SNIa + 25 elliptical $\beta$ model sample, and Union 2.1 SNIa + 38 spherical $\beta$ model sample, respectively. For the SNIa data, only the statistical errors are considered. The shaded region shows the constraint from the latest Planck measurement with the $1\sigma$ CL ($H_0=67.8\pm 0.9 \; {\rm km\; s^{-1}\; Mpc^{-1}}$)~\cite{Planck XIII}. \label{Fig3}}
\end{figure}

In this case, we consider the Union 2.1 SNIa data without and with the systematic error. For the latter case, the covariance of the SNIa data in a bin is considered using Eqs.~(\ref{bu2}) and (\ref{sig2}). Because the uncertainty of the galaxy cluster data is much larger than that of the SNIa data, the systematic error of the SNIa data has a negligible effect on the results. For example, at the $1\sigma$ CL, $h=0.5940\pm 0.0303$ and $h=0.5938\pm 0.0303$ for 25 galaxy clusters + Union 2.1 SNIa without and with systematic errors, and $h=0.6395\pm 0.0298$ and $h=0.6393\pm 0.0298$ for 38 galaxy clusters + Union 2.1 SNIa without and with systematic errors, respectively. Therefore, we plot only the likelihood distributions of $h$ from galaxy clusters + Union 2.1 SNIa without the systematic errors, which are shown in Fig.~(\ref{Fig3}), where the blue and purple solid lines represent the results from the Union 2.1 SNIa + elliptical $\beta$ model sample and Union 2.1 SNIa + spherical $\beta$ model sample, respectively. The shaded region shows the constraint from the latest Planck measurement at the $1\sigma$ CL ($H_0=67.8\pm 0.9 \; {\rm km\; s^{-1}\; Mpc^{-1}}$)~\cite{Planck XIII}. 

Comparing Figs.~(\ref{fig2}) and (\ref{Fig3}) reveals that, for the same galaxy cluster sample, different SNIa samples give almost the same result, but Union 2.1 favors a slightly larger $h$. The combinations of SNIa and the elliptical $\beta$ model sample seem to favor a smaller Hubble constant, which is marginally consistent with the latest Planck result at the $3\sigma$ CL. The Hubble constants from SNIa + spherical $\beta$ model are compatible with that from Planck at the $2 \sigma$ CL and are quite consistent with that reconstructed from $H(z)$ data directly given in \cite{Busti2014}, where $H_0=64.9\pm 4.2 \; {\rm km\; s^{-1}\; Mpc^{-1}}$. It is easy to see that different galaxy cluster samples give different $h$ constraints. This finding is the same as that obtained in \cite{Holanda2014} on the basis of a $\Lambda$CDM model in which galaxy cluster samples were combined with other data. However, our values for the Hubble constant are much less than those given in \cite{Holanda2014}. Thus, there is a slight incompatibility between the galaxy cluster + SNIa and Planck results, which is the same as that obtained in~\cite{Planck}, where some inconsistencies between the cosmological parameter constraints derived from the Planck SZ cluster measurements and from the Planck CMB data were found. This suggests that assumptions about global cluster properties strongly affect the cosmological model inferred.
\section{baryon acoustic oscillations}
BAOs (see \cite{Bassett09} for a review) arise from the coupling of photons and baryons by Thomson scattering in the early universe. Competition between the radiation pressure and gravity gives rise to a system of standing sound waves within the plasma. At recombination, the interaction between photons and baryons ends abruptly because the free electrons are quickly captured, which leads to a slight overdensity of baryons at the sound horizon. The corresponding scale is the distance traveled by a sound wave in the plasma before recombination. This scale has been measured in the clustering distribution of galaxies today and can be used as a standard ruler. Combining measurements of the baryon acoustic peak and the Alcock--Paczynski distortion from galaxy clusters, the ADD data can be obtained. Note that, to obtain these data, a fiducial cosmological model is still needed. We summarize five low-redshift data points in Table~(\ref{Tab1}). Three of them are determined from the WiggleZ Dark Energy Survey~\cite{Blake2012}, and the other two data points are from BOSS DR7~\cite{Xu2013} and DR11~\cite{Samushia2013}.
\begin{table}
 \begin{center}
\begin{tabular}{c c c}
  \hline\hline
        $z$ &   $D_A(z)$ (Mpc)      &   Survey \\  \hline
   0.44 &     $1205\pm114$    &   WiggleZ\cite{Blake2012}       \\
    0.6 &     $1380\pm95$    &         \\
     0.73 &     $1534\pm107$    &        \\   \hline
 0.35 &    $ 1050 \pm38$    &     SDSS DR7\cite{Xu2013}   \\   \hline
   0.57 &    $ 1380 \pm23$    &     SDSS DR11 CMASS\cite{Samushia2013}   \\   \hline
   \hline
   \end{tabular}
\tabcolsep 3pt \caption{\label{Tab1} Summary of the ADD measurements from the BAOs.  }
    \vspace*{-12pt}
       \end{center}
       \end{table}
\begin{figure}
\includegraphics[width=8cm]{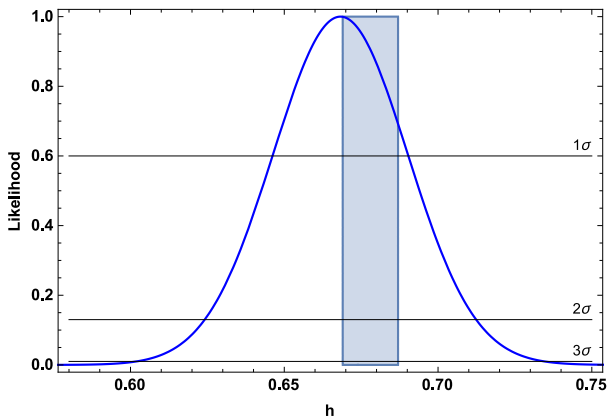} 
\caption{The likelihood distributions of $h$ from BAOs + SDSS-II SNIa. The shaded region shows the constraint from the latest Planck measurement with the $1\sigma$ CL ($H_0=67.8\pm 0.9 \; {\rm km\; s^{-1}\; Mpc^{-1}}$)~\cite{Planck XIII}. \label{Fig4}}
\end{figure}
\begin{figure}
\includegraphics[width=8cm]{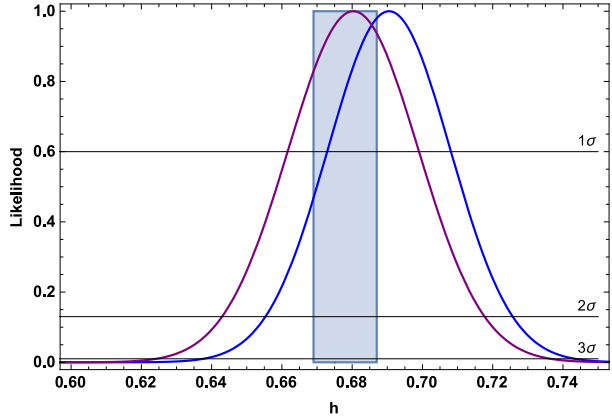} 
\caption{The likelihood distributions of $h$ from BAOs and Union 2.1 SNIa data. Blue and purple lines represent the results without and with the systematic errors of SNIa considered in a bin. The shaded region shows the constraint from the latest Planck measurement with the $1\sigma$ CL ($H_0=67.8\pm 0.9 \; {\rm km\; s^{-1}\; Mpc^{-1}}$). \label{Fig5}}
\end{figure}
\subsection{BAOs + SDSS-II SNIa}
The likelihood distribution of $h$ from BAOs + SDSS-II SNIa is shown in Fig.~(\ref{Fig4}). As in the previous section, only the statistical errors are considered for SDSS-II SNIa. At the $1\sigma$ CL, we have $h=0.6683\pm 0.0221$, which is tighter than the values from the galaxy cluster samples, although only five BAO data points are used. It is more consistent with the result from Planck than the values from the galaxy cluster samples. This consistency occurs at the $1\sigma$ CL.
\subsection{BAOs + Union 2.1 SNIa}
Fig.~(\ref{Fig5}) shows the likelihood distributions of $h$ from BAOs and Union 2.1 SNIa. Blue and purple lines represent the results without and with the systematic errors of SNIa considered in a bin, respectively. A strong constraint on $h$ is clearly obtained: $h=0.6905\pm 0.0176$ for BAOs + SDSS-II SNIa without the systematic errors, and $h=0.6803\pm0.0187$ for BAOs + SDSS-II SNIa with the systematic errors at the $1\sigma$ CL, an approximately $2.5\%$ determination. These results are consistent with those from Planck at the $1\sigma$ CL, and the consistency improves when the systematic errors in a bin are included. 

The SDSS-II SNIa clearly favor a slightly smaller $h$, which is the same as that for the galaxy clusters. The values from SNIa + BAOs are larger than those from SNIa + galaxy cluster samples. They are quite consistent with the results from Planck with and without the systematic errors of the $217\times 217 \; {\rm GHz}$ detector set spectrum considered~\cite{Planck, Planck XIII, Spergel 2013} and the combination of all BAO data~\cite{Cheng2014}. It is also compatible with the local measurement from Cepheids and very-low-redshift SNIa given in \cite{Efstathiou2014, Rigault14}. 
\section{conclusions}
In this paper, we propose a model-independent method to determine the value of the Hubble constant, which is a very important constant in cosmology. Because the distance modulus of SNIa depends on the chosen value of $h$, we obtain an $h$-dependent expression for the LD, which is given in Eq.~(\ref{DL}). Thus, if the LD at a redshift can be determined by other observations, it can be combined with the one from SNIa to determine the value of $h$ using the fact that two different observations should give the same value of the LD at a given redshift. According to the distance duality relation, the LD can be deduced from the observed ADD data. Using the SDSS-II SNIa released by the MLCS2k2 light curve fitter to provide the observed LD data and the galaxy cluster samples to provide the observed ADD data, we find that $h=0.5867\pm0.0303$ for the elliptical $\beta$ model sample, and $h=0.6199\pm0.0293$ for the spherical $\beta$ model sample. The former is smaller than the values from other observations, whereas the latter agrees very well with the value directly reconstructed from the $H(z)$ data~\cite{Busti2014}. Thus, different galaxy cluster samples give different values of $h$. This result is the same as that obtained in ~\cite{Holanda2014} on the basis of a $\Lambda$CDM model with a combination of galaxy clusters and other observations. If the ADD data from BAO measurements are considered, we obtain $h=0.6683\pm 0.0221$ at the $1\sigma$ CL, which is larger than the values from SNIa + galaxy clusters. In addition, for a comparison, we consider the Union 2.1 SNIa sample given by the SALT2 light curve fitter. The results from Union 2.1 SNIa are model-dependent, as the distance modulus from SALT2 is determined in the framework of the $\Lambda$CDM model. The Union 2.1 SNIa give a slightly larger value of $h$ than the SDSS-II one. The covariance of the Union 2.1 data has an apparent effect on the result when BAO data are used. We find that the results from SNIa + BAOs are quite consistent with those from Planck~\cite{Planck, Planck XIII, Spergel 2013} and the combination of all BAO data~\cite{Cheng2014}, as well as the local measurement of Cepheids and very-low-redshift SNIa~\cite{Efstathiou2014, Rigault14}.
\acknowledgments
We thank  Prof. Shuang-Nan Zhang  and Prof. Xiao-Feng Wang  for their useful discussions. This work was supported by the National Natural Science Foundation of China under Grants No. 11175093, No. 11222545, No. 11435006, and No. 11375092;  the  Specialized Research Fund for the Doctoral Program of Higher Education under Grant No. 20124306110001;  and the K.C. Wong  Magna Fund of Ningbo University.

 \end{document}